\documentclass[12pt]{iopart}
\usepackage{graphicx}
\usepackage{pstricks,pst-node}
\usepackage{pst-node,pst-text,pst-3d} 
\usepackage{verbatim}
\usepackage{tabularx}
\relax 
\input epsf.tex

\font\af=msbm12

\def\C{{\af C}}

\def\CC{{\cal C}}

\def\CM{{\cal M}}
\def\CU{{\cal U}}
\def\CU{{\cal U}}
\def\sp #1{{{\cal #1}}}
\def\be{\begin{eqnarray}}    
\def\ee{\end{eqnarray}}
\def\Dsl{\,\raise.05ex\hbox{/}\mkern-9.5mu D}
\def\mbox#1#2{\vcenter{\hrule \hbox{\vrule height#2in 
\kern#1in \vrule} \hrule}} 
\def\boxeqn#1{\vcenter{\vbox{  \hrule height2pt \hbox{\vrule
width 2pt \kern3pt\vbox{\kern3pt
\hbox{${\displaystyle #1}$}\kern3pt}\kern3pt\vrule width 2pt}\hrule height2pt}}}

\def\back{{{\raise.4em\hbox{$\, _\backslash\,$}}}}

 2

\font\blackboard=msbm10 \font\blackboards=msbm7
\font\blackboardss=msbm5
\newfam\black
\textfont\black=\blackboard
\scriptfont\black=\blackboards
\scriptscriptfont\black=\blackboardss

\let\spec=\blackb
\def\sp #1{{{\cal #1}}}

\def\sp #1{{{\cal #1}}}

\def\frac#1#2{{#1\over #2}}

\def\big R{{\hbox{{\bigfield R}}}}
\def\bbig R{{\hbox{{\bbigfield R}}}}

\font\af=msbm10

\font\afm=msbm12

\def\Z{\hbox{\af Z}}

\def\C{\hbox{\afm C}}

\def\Zm{\hbox{\afm Z}}
\def\I{\hbox{\afm I}}

\mathchardef\imath="717B
\def\inbar{\,\vrule height1.5ex width.4pt depth0pt}
\def\IB{\relax{\rm I\kern-.18em B}}
\def\IC{\relax\hbox{$\inbar\kern-.3em{\rm C}$}}
\def\ID{\relax{\rm I\kern-.18em D}}
\def\IE{\relax{\rm I\kern-.18em E}}
\def\IF{\relax{\rm I\kern-.18em F}}
\def\IG{\relax\hbox{$\inbar\kern-.3em{\rm G}$}}
\def\IH{\relax{\rm I\kern-.18em H}}
\def\II{\relax{\rm I\kern-.18em I}}
\def\IK{\relax{\rm I\kern-.18em K}}
\def\IL{\relax{\rm I\kern-.18em L}}
\def\IM{\relax{\rm I\kern-.18em M}}
\def\IN{\relax{\rm I\kern-.18em N}}
\def\IO{\relax\hbox{$\inbar\kern-.3em{\rm O}$}}
\def\IP{\relax{\rm I\kern-.18em P}}
\def\IQ{\relax\hbox{$\inbar\kern-.3em{\rm Q}$}}
\def\IR{\relax{\rm I\kern-.18em R}}
\font\cmss=cmss10 \font\cmsss=cmss10 at 10truept
\def\IZ{\relax\ifmmode\mathchoice
{\hbox{\cmss Z\kern-.4em Z}}{\hbox{\cmss Z\kern-.4em Z}}
{\lower.9pt\hbox{\cmsss Z\kern-.36em Z}}
{\lower1.2pt\hbox{\cmsss Z\kern-.36em Z}}\else{\cmss Z\kern-.4em Z}\fi}
\def\IGa{\relax\hbox{${\rm I}\kern-.18em\Gamma$}}
\def\IPi{\relax\hbox{${\rm I}\kern-.18em\Pi$}}
\def\ITh{\relax\hbox{$\inbar\kern-.3em\Theta$}}
\def\IOm{\relax\hbox{$\inbar\kern-3.00pt\Omega$}}

\def\CC{{\cal C}}

\def\CM{{\cal M}}

\def\CH{{\cal H}}

\def\CU{{\cal U}}

\begin{document}

\title{Casimir Effect and Global Theory of Boundary Conditions}

\author{M. Asorey, D. Garc\'\i a \'Alvarez, J. M.  Mu\~noz-Casta\~neda}

\address{Departamento de F\'\i sica Te\'orica. Facultad de Ciencias.
Universidad de Zaragoza, 50009 Zaragoza.Spain}
\begin{abstract}

The consistency of quantum field theories  defined on domains with external 
borders  imposes very restrictive constraints 
on the type of boundary conditions that the fields can satisfy.
We analyse the global geometrical  and topological properties of the space of
all possible boundary conditions for scalar quantum field theories.
The variation of the Casimir energy under the change of
boundary conditions reveals the existence of singularities
generically associated to boundary conditions which either
involve topology changes of the underlying physical space or 
edge states with unbounded below classical energy.  
The effect can  be understood in terms of
a new type of  Maslov index associated to the non-trivial
topology of the space of boundary conditions.  We also
analyze the global aspects of the 
renormalization group flow, T-duality and the conformal invariance of the
corresponding fixed points. 

\end{abstract}

\maketitle

\section{Introduction}
The role of boundaries in Quantum Physics has being boosted in the last decade
until  becoming  a basic element of fundamental physics. 
The classical Weyl's  dream of hearing the shape of a drum has been subsumed into a quantum
dream of hearing the shape of a quantum drum or even a more dramatic one that of the shape of our
Universe. 


 In quantum mechanics  the unitarity principle,
imposes severe constraints on the boundary behaviour of quantum
states in systems restricted to bounded domains \cite{aim}. In relativistic field theories,
causality imposes further requirements \cite{gift}. The space of boundary conditions
compatible with both constraints  has interesting global geometric properties.
The dependence of many  interesting physical phenomena, like the Casimir effect \cite{11},
topology change \cite{14}  or  renormalization group flows
\cite{Affleck,Zuber,Cardy}, on the boundary conditions can be  analyzed from this global perspective.

The effect of  background fields in  quantum theories has  been extensively
analyzed from many perspectives. The induced dynamics on the background field
by the effective action has many interesting implications \cite{Saviddy,Abbott}. The analogue study
with respect to possible boundary conditions has not been yet globally addressed,
and it is the main purpose of this paper.

\section{Selfadjointness  boundary conditions}

Let us for simplicity consider $N$ massless free complex scalar fields 
$\phi$  defined on a bounded
domain  $  \Omega$ with smooth boundary $ \partial\Omega$. The corresponding Hamiltonian
is given by
\be
{\cal H}=\frac{1}{2}\int_\Omega [|\pi_\phi|^2+|\nabla \phi |^2+m^2|\phi|^2]-\frac1{4}\int_{\partial\Omega} \left[\phi^\dagger  \,
\partial_n{\phi} -
(\partial_n{\phi^\dagger})\phi\right]  , 
\label{one}
\ee
where  the boundary term is introduced to generate local classical equations of 
motion equation without requiring any specific type of   boundary conditions \cite{bfsv,saharian}.
Indeed, the gradient term
\be
{\cal V}=\frac{1}{2}\int_\Omega |\nabla \phi |^2
\ee
can be rewritten as
 \be
{\cal V}=\frac{1}{2}\int_\Omega \phi^\dagger  \Delta \phi+\frac{1}{2}\int_{\partial\Omega} \phi^\dagger  \, \partial_n \phi
\ee
where $\partial_n $ denotes the normal derivative at the boundary $\partial\Omega$.
If the space has a non-trivial   Riemanian metric $g$ the volume elements
in the domain $\Omega$ and its boundary $\partial \Omega $ are
\be
\delta \,\Omega =    \sqrt{g}\, d^nx;             \quad  {\rm and}\quad    \delta \, \partial\Omega  =  {\ \sqrt{g_{_{\partial\Omega}}}d^{{n-1}}x},
\ee
respectively. $ \Delta$  is the Laplace-Beltami operator $ \Delta =-{\  \,\nabla^\mu}\partial_{\nu} $ .

In the classical field theory, boundary conditions have to be imposed on the fields 
in order to find a unique solution of motion equations.
In the quantum theory, boundary conditions have to be imposed in order to
preserve unitarity.
In particular, the Laplace-Beltami operator must  be a selfadjoint
operator. The standard theory of self-adjoint extensions due to von Neumann \cite{vn}
establishes that  there exists an one-to-one correspondence between
self-adjoint extensions of $\  \Delta$ and unitary operators  from the
deficiency spaces 
$\ \sp N_+ = \ker (\Delta^\dagger  + i\, \I)$ to $\ \sp N_- = \ker (\Delta^\dagger  - i\, \I)$.
There is, however,  an alternative characterization \cite{aim} based on explicit 
constrains on  boundary data which is more practical for physical applications. It
establishes that the set $\  \CM$ of self-adjoint extensions of
$\  \Delta$ is in one-to-one correspondence with the group of unitary
operators of  the boundary Hilbert space $\  L^2(\partial\Omega,\C^N)$.  For
 any  unitary operator $U\in{\CU} (L^2(\partial\Omega,\C^N))$, the fields satisfying the 
boundary condition \footnote{Notice that  (\ref{const})
 is in general a non-local condition. U is 
any unitary operator of $L^2(\partial\Omega,{\hbox{\af C}}^N)$ that generically
    is non-local.}
\be
{\phantom{\Bigl[}  
\varphi- i \partial_n\varphi ={   U} \left(\varphi+ i \partial_n\varphi
\right)\phantom{\Bigr[} }
\label{const}
\ee
define a domain where $\  \Delta$ is a selfadjoint operator. $\varphi$ 
denotes the boundary value of $\phi$ and  $\partial_n\varphi$ its normal derivative at the 
boundary.  
Although, both characterizations are equivalent \cite{aim}, the later
provides a group structure to the space $\CM$ of boundary conditions and
allows a more direct analysis of its global properties.

In the case of open strings, for the corresponding conformal  1+1 dimensional theories 
defined on the space interval  $\Omega =[0,1]\subset \IR$ we have $\CM=U(2)$.
The unitary matrices
\be
\hspace{-2cm}
U_D=-\I= \begin{pmatrix}
{{-1}&{0}\cr
{0}&{-1}} 
\end{pmatrix}\qquad 
U_N=\I=\begin{pmatrix}{{1}&{0}\cr
{0}&{1}}\end{pmatrix}\qquad 
U_P=\sigma_x=\begin{pmatrix}{{0}&{1}\cr
{1}&{0}}\end{pmatrix}
\ee
define  Dirichlet, Neumann 
and  periodic  boundary conditions, which in string theory correspond to
a string  attached to a
D-brane background, free  open  and closed string theories. respectively.

For  higher N-dimensional target spaces, or N-component strings, 
the  2N$\times$2N  matrices 
\be
\hspace{-2cm}
U_1=\begin{pmatrix}{
{0}&{0}&{0}&{0}&{0} &{\cdots}&{0} & {1}\cr
{0}&{0}&{1}&{0}&{0} &{\cdots } & {0}&{ 0}\cr
{0}&{1}&{0}&{0}&{0} &{\cdots } & {0}&{ 0}\cr
{0}&{0}&{0}&{0}&{1} &{\cdots} &{0}&{ 0}\cr
{0}&{0}&{0}&{1}&{0} &{\cdots} &{0}&{ 0}\cr
{\cdot} &{\cdot}&{\cdot}&{\cdot}&{\cdot}&{\cdots}&{ \cdot}&{ \cdot}\cr
{1}&{0}&{0}&{0}&{0}&{\cdots}&{0}&{ 0}\cr}
\end{pmatrix}\quad
U_N=\begin{pmatrix}{
{0}&{1}&{0}&{0}&{0} &{\cdots}&{0} & {0}\cr
{1}&{0}&{0}&{0}&{0} &{\cdots } & {0}&{ 0}\cr
{0}&{0}&{0}&{1}&{0} &{\cdots } & {0}&{ 0}\cr
{0}&{0}&{1}&{0}&{0} &{\cdots} &{0}&{ 0}\cr
{\cdot} &{\cdot}&{\cdot}&{\cdot}&{\cdot}&{\cdots}&{ \cdot}&{ \cdot}\cr
{0}&{0}&{0}&{0}&{0}&{\cdots}&{0}& {1}\cr
{0}&{0}&{0}&{0}&{0}&{\cdots}&{1}&{ 0}\cr}
\end{pmatrix}
\ee
define selfadjoint extensions which correspond to one single closed string 
or  N  disconnected strings, respectively. The topology change is described
in this way  by
a simple change of boundary conditions in $\CM$.


\section{Global topological structure  of the space of selfadjoint boundary conditions }
There are subsets of boundary conditions in $\CM$
where the constraint (\ref{const}) acquires a simpler expression.
If the spectrum of eigenvalues of the unitary operator $U$ does not include
the value $-1$ (i.e.
$\  -1\notin{\rm Sp}\, U$) the boundary condition (\ref{const}) can be rewritten
as 
\be 
\partial_n \varphi=-i {\I-U\over \I+U} \varphi
\ee
 which means that only the boundary value of the fields at the boundary can
 have an arbitrary value  $\varphi $ 
whereas its normal derivative is determined by $U$ and $\varphi $.

The corresponding operator mapping  from unitary into selfadjoint operators 
\be  
A=-i {\I-U\over \I+U}
\ee
is the celebrated Cayley transform. The inverse Cayley transform
\be
 U= {\I-iA\over \I +iA}
\ee
recovers the unitary operator $U$ from its selfadjoint Cayley transform $A$.

 If $\  -1\notin{\rm Sp}\, U$  we can interchange the role of $\varphi $ and  $ \partial_n \varphi$. In
 that case the boundary condition reads 
\be
 \varphi=i {\I+U\over \I-U} \partial_n\varphi.
\ee

However, there are two submanifolds ({\it Cayley submanifolds})  of $\CM$ defined by 
\be  {\sp C_\pm} =  \left\{U\in \sp U\left( L^2 (\partial\Omega,\C^N)\right)\Big|  \pm 1 \in  {\rm Sp}\,  U\right\}\ee
 where both transformations are singular.

The topology of the space $\CM$ of selfadjoint extensions is non-trivial 
\be
 \pi_1\,\left[\CU(\  L^2(\partial\Omega,\C^N))\right]=\Zm
\ee
and the Cayley submanifolds are homologically dual of the generating cycles of
$H_1(\CM)$ \cite{aim}. A generalized  Maslov  index 
can be defined for  any closed path $\ \gamma\in \CM $ as  the 
oriented sum of crossings of $\  \gamma$ across the Cayley submanifold $\ \sp C_-$, i.e.
\be
  \nu_M(\gamma)=\int_0^{2\pi}\partial_\theta n (\gamma(\theta) ) d\theta.
\ee
where $n\left(\gamma(\theta)\right)=n_+\left(\gamma(\theta)\right) -n_-\left(\gamma(\theta)\right)$ denotes the indexed sum 
of crossings of $\left(\gamma(\theta')\right)$ for $\theta'\leq \theta $.
A relevant consequence of  the non-trivial structure of the
space of boundary conditions is that  a Berry phase can appear when the system
follows a
non-trivial is closed loop  in the space of boundary conditions.

Boundary conditions which correspond to the identification of points of the boundary
can easily be identified because their unitary matrices $U$ present pairs of
eigenvalues $\pm 1$. The unitary operators associated  to these boundary
conditions  belong to  the intersection of the Cayley manifolds $\CC_+$ and $\CC_-$.
The transition from  normal boundary conditions to any of these  conditions involves
a topology change. Now,  such a topological transition always
requires  an infinite amount of  classical energy \cite {aim}.
This property follows from the fact that  for any selfadjoint extension  of  $\Delta$ with
$ U\in \CC_-$, there exist a  family of selfadjoint  extensions with unitary
operators  $U_t$  very close to $U$ that have bounded   edge states 
with  negative classical energy   $E_{-}$   which 
diverges in the limit $U_t\to U$.
 
Classical  fields  with negative energy are possible if the the Cayley transform operator $A$ is
not negative, because
\be  \int_\Omega \phi^\dagger   \Delta \phi  = \int_\Omega |  \nabla\, \phi|^2 - \int_{\partial\Omega}\varphi^\ast  A
\varphi.\ee
However, non-positive  selfadjoint extensions of $\Delta$ might lead to inconsistencies
in the quantum field theory if they are not bounded below by the mass term.

\section{Casimir energy and boundary conditions}

The infrared properties of quantum field theory are very sensitive to boundary
conditions \cite{karpacz}. In particular, the physical properties of the quantum 
vacuum state and the vacuum energy exhibit  a very strong dependence on 
the type of boundary conditions.  Let us consider, for simplicity, the case of a real massless
field in 1+1 dimensions defined on a finite interval $[0,L]$.

For  pseudoperiodic boundary conditions defined by the unitary operator
\be
  U_\theta =\cos  \theta\, {\  \sigma_x} - \sin \theta \, \  \sigma_y \qquad  \varphi(L)={\rm e}^{i \theta} \varphi(0)
\label{pseudo}
\ee
the Casimir vacuum energy (see e.g. Ref. \cite{bordag} and references
therein) is given by (Fig. 1)
\be  E_0=\frac{\pi}{
  L}\left(\frac{1}{12}-{\min_{n\in\Z}}\left({ \frac{\theta}{2\pi}\,}{+n}-\frac1{2}\right)^2\right)
\ee
\centerline{\  \epsfxsize8cm
 \epsfbox{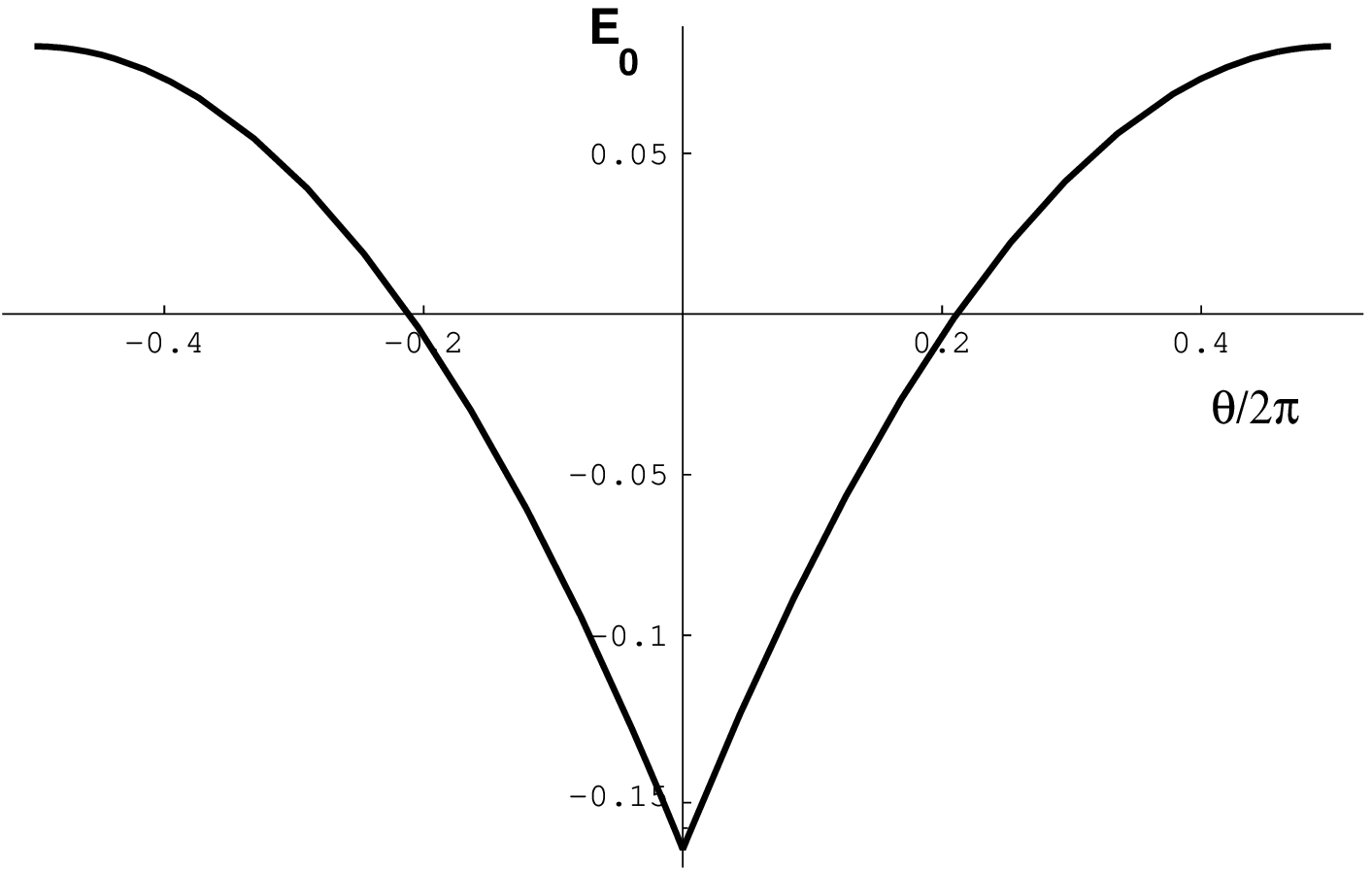}
}
\centerline{\small {\bf Figure 1.} Casimir Energy for pseudo-periodic boundary conditions}
\vspace{.0cm}

The vacuum energy dependence is in this case is relatively smooth. It 
presents a cuspidal point at $\theta=0$ which corresponds to periodic
boundary conditions. A completely regular  behaviour is obtained
for Robin boundary conditions (Fig. 2)
\be
  U={\rm e}^{2 \alpha i}\I \, \quad  \partial_n{\varphi}(0)=\tan \alpha{}\, \varphi(0),\
\partial_n{\varphi}(L)={\tan}\, \alpha{}\, \varphi(L) 
\label{robs}
\ee
which smoothly  interpolate between Dirichlet ($\alpha=\frac{\pi}{2}$) and Neumann ($\alpha
=\pi$) boundary conditions when $\alpha$ is restricted to  the interval  $\alpha\in
[\frac\pi{2},\pi]$ \cite{aa, cavalcanti, farina} .
\centerline{\  \epsfxsize8cm\epsfbox{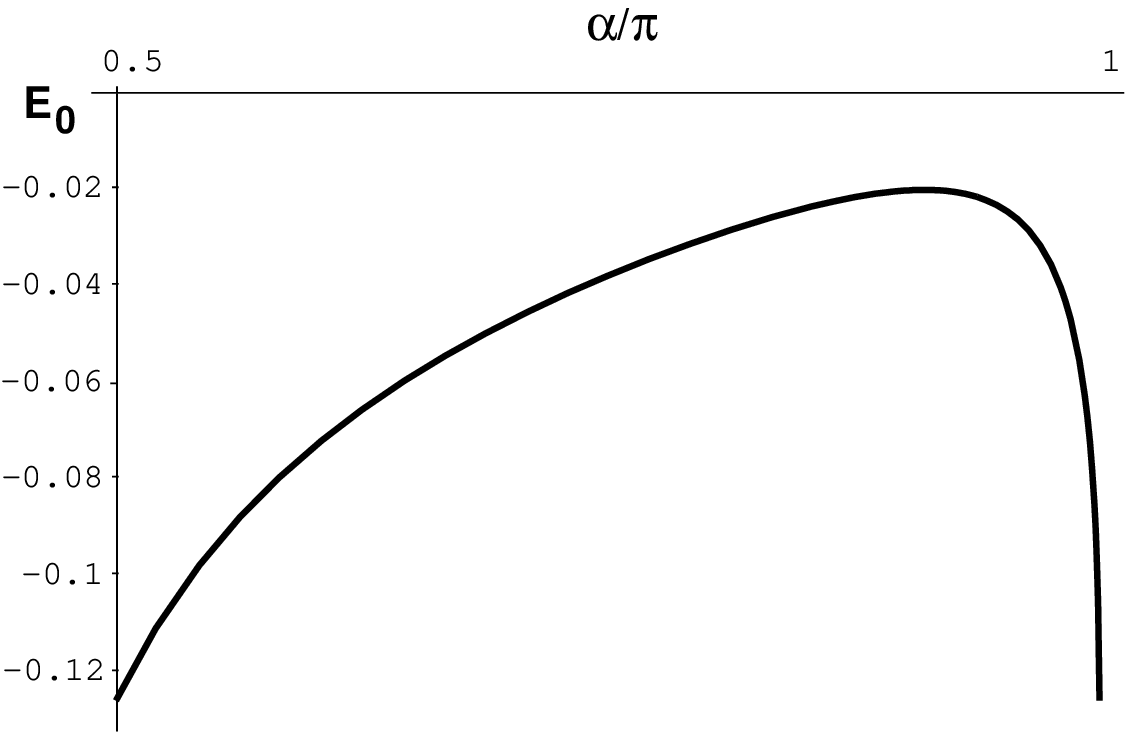}}

\centerline{\small {\bf Figure 2.}  Vacuum Energy for Robin boundary conditions}
\medskip

However, the vacuum energy can exhibit more singular 
behaviours when globally considered as  function  defined  on $\CM$.
Indeed, for  Robin boundary conditions (\ref{robs}) in the interval
$0<\alpha< \frac{\pi}{2}$ the Casimir energy acquires an imaginary contribution
due to the appearance of negative classical energy modes associated to  edge
states (Fig. 3). 
\medskip

\centerline{\  \epsfxsize8cm
 \epsfbox{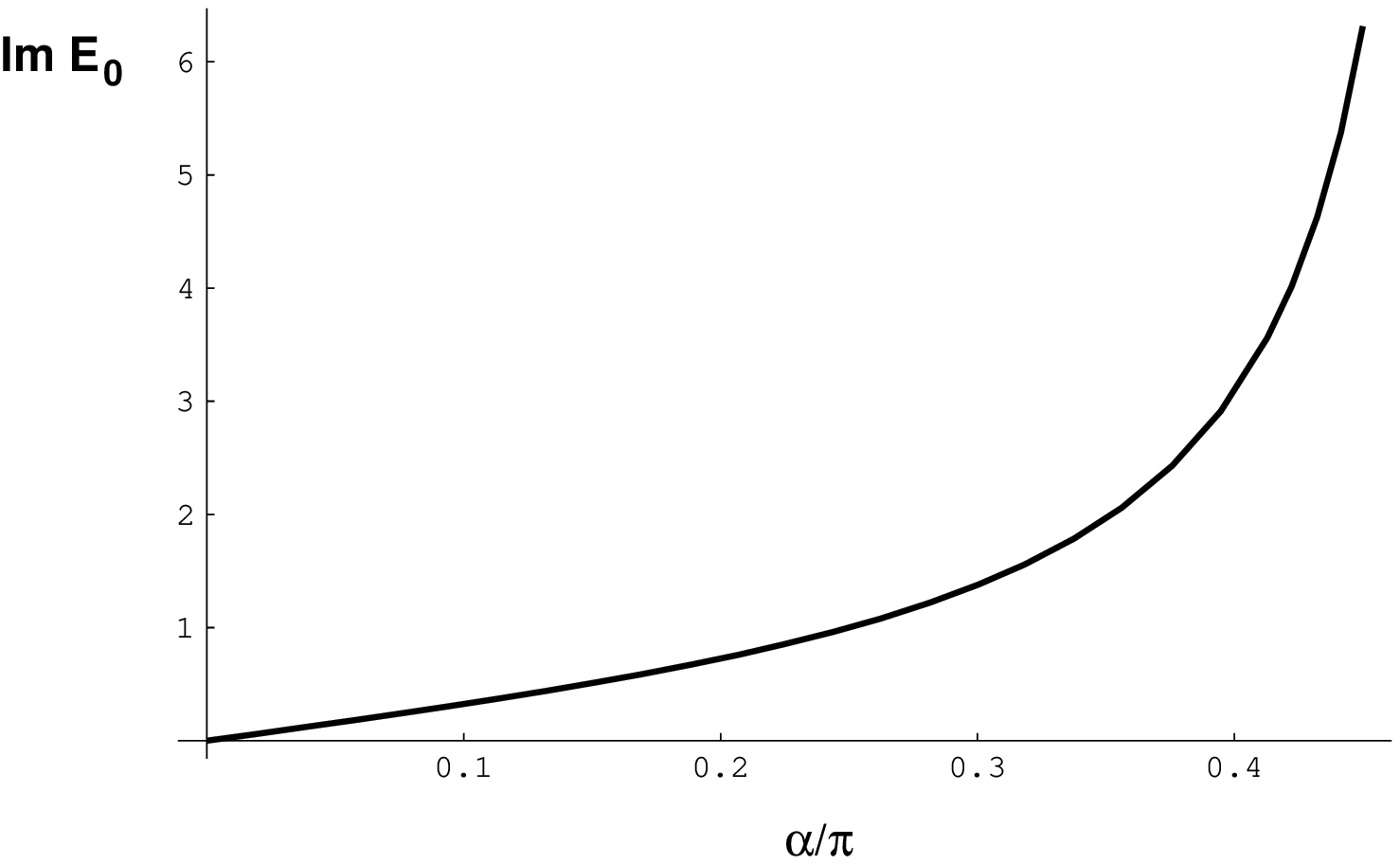}
}
{\hbox{\small {\bf Figure 3.}  Behaviour of the imaginary term of the
Casimir energy for  Robin boundary }}
\hspace{1cm}\hbox{conditions}
\medskip

\noindent
The fact that the classical energy of this edge state becomes unbounded below when
$\alpha\to \frac\pi{2}$ implies a pathological behaviour of  Casimir energy around 
the Dirichlet boundary condition point. 

The existence of edge states giving rise to complex Casimir energies 
is a  generic feature. For instance, the phenomenon
also appears  for Robin boundary conditions of the type
\be
 U=\begin{pmatrix} {{{\rm e}^{2\alpha i}}&{0}\cr
{0}&{{\rm e}^{-2\alpha i }}}
\end{pmatrix}
\, \quad  \partial_n{\varphi}(0)={\tan \alpha}{\,}\varphi(0),\ 
\partial_n{\varphi}(L)=-{\tan \alpha}{\,}\varphi(L) 
\ee
except when $\alpha=n\pi/2$ which again correspond to Neumann and Dirichlet
conditions \cite{caxambu}.
In all the cases  with imaginary vacuum energy
the Hamiltonian 
\be
\CH=\frac1{2}\sqrt{\Delta^U}
\label{hamilton}
\ee
is not even a selfadjoint operator  due to the existence
of an edge estate  of the selfadjoint operator $\Delta^U$ with negative eigenvalue.
The appearance of an imaginary part in the Casimir energy  can be associated 
to a pair creation phenomenon. In all  the above  cases the imaginary term also
becomes   
singular in the limit $\alpha \to 0$ as a consequence of the AIM theorem 
\cite{aim}. 
However, the borderline regime $\alpha=\frac\pi{2}$ is  always very regular as corresponds to
Dirichlet  boundary conditions. The same phenomenon occurs around any boundary
condition involving a topology change.



\section{  Boundary Conditions and Symmetries}
\medskip

The consistency of the quantum field theory imposes, thus, a very
stringent condition on the type of acceptable boundary conditions even in the
case of massive theories in order to prevent this type of pathological behaviour of vacuum energy.

Moreover, because of the existence of the boundary term in (\ref{one})
 the Hamiltonian $\CH$ (\ref{hamilton}) is not selfadjoint  if the spectrum of the unitary
operator $U$ intersects the following domain of phase factors  
$$
S^1_{m}=\{{\rm e}^{2\alpha i}; -\pi <\alpha \leq \pi ,0<\alpha<\frac{\pi}{2}- \,\arctan\, {m^2}, {\rm
  or}\, \frac\pi{2}<-\alpha<{\pi}- \,\arctan\, {m^2}\, \}.
$$
In any other case, 
 $-m^2$ is a lower bound
for the spectrum of the operator $\Delta^U$ and $\CH$ is selfadjoint.

The  space of  consistent boundary conditions for the quantum
field theory
\be
\CM_{m}=\{U\in \CU(\  L^2(\partial\Omega,\C^N));\ {\rm Sp}\, U\cap S^1_{m} = \emptyset \}
\ee
is not necessarily multiple-connected which means that can have no
Maslov index, although $\CM_m$
might also  have a non-trivial topology.

For real scalar fields there is a further condition. $U$ has to satisfy
a  CP  symmetry preserving condition
 \be
U^\dagger=U^\ast ,\quad U=U^T .  
\ee
The usual Neumann and Dirichlet boundary conditions
${  U=\pm \I \  }$ satisfy this requirement.
In general, for
\be
\  U= \begin{pmatrix}{
{A_1}&{B}\cr
{B^T}&{A_2} }
\end{pmatrix}
\ee
the condition requires that
\be
 A_1=A^T_1,\, A_2=A^T_2 ,\, A_1 B^\ast +B A_2^\dagger =0\ee
\be
 B B^\dagger   +A_1A_1^\dagger =\I, \, A_2A_2^\dagger +B^TB^\ast=\I 
\ee

In particular, the quasi-periodic condition
${\varphi}(L)= M^{-1}{\varphi}(0),\ 
 \partial_n{\varphi}(L)= -M \partial_n{\varphi}(0)$ is also compatible if $ M=M^t=M^\ast$.

In the case of a single  real massless scalar the set of compatible
boundary conditions is reduced from $\CM=S^3$ down to  $\CM=\Zm_{_2}{{\spec{n}}} S^1$
which also has a group structure and two connected components: $\CM_0$
 given by the operators of the form ${  U_{\pm}=\pm\, \I }$
and $\CM_1$ given by 
\be U_{\alpha}=\cos\alpha\,{ \  \sigma_z} + \sin \alpha\,
  {\  \sigma_x}\, .
\label{quasi}
\ee
 $\CM_0$  includes Neumann  and Dirichlet conditions; and $\CM_1$
contains the  quasi-periodic boundary conditions 
\be
\varphi(L)={\tan \frac\alpha{2}}\,\varphi(0);\qquad \partial_n{\varphi}(L)=-\left({\tan \frac\alpha{2}}\right)^{-1}{\,}\partial_n{\varphi}(0)
\ee
which include periodic ($\alpha=\frac\pi{2}$)  and antiperiodic ($\alpha=-\frac{\pi}{2}$)  boundary
conditions.

In this case the topology of the global set of boundary conditions is not
connected $\pi_0(\CM)=\Zm_{2}$ and has a Maslov index, $\pi_1(\CM_1)= \Zm$. 

$\,$\vspace{.1cm}

\centerline{\  \epsfxsize8cm \epsfbox{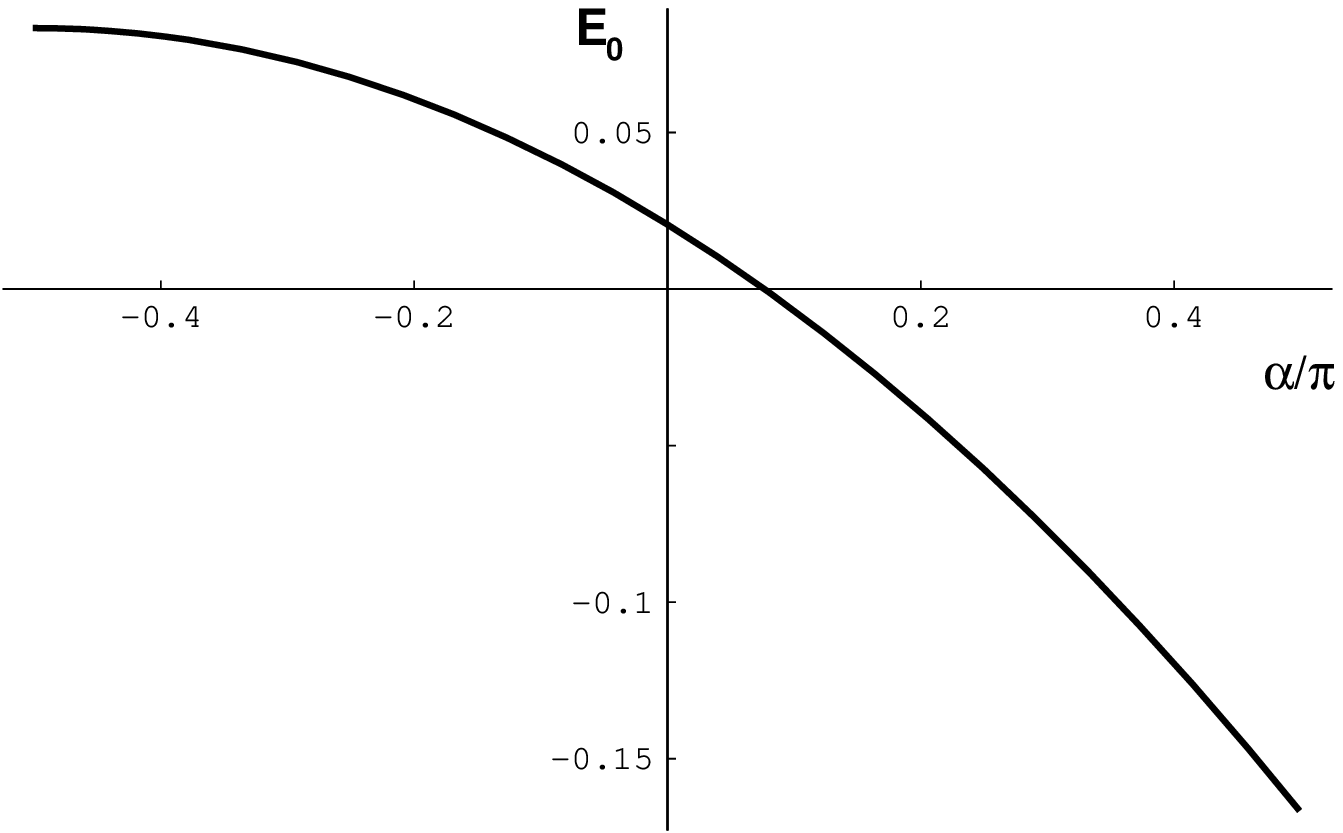}}

\centerline{\small {\bf Figure 4.}  Casimir energy for  quasi-periodic
  boundary conditions}
\medskip

The   Casimir energy for quasi-periodic boundary conditions is \cite{gift}
\be
 E_0=\frac{\pi}{
  L}\left(\frac{1}{12}-{\min_{n\in\Z}}\left(\frac{  \alpha}{2\pi}+n+\frac1{4}\right)^2\right)
\label{qp}
\ee
(see figure 4).

Two particularly  interesting cases of quasi-periodic boundary conditions are
given by $\alpha=0$, 
\be\  U_Z={\  \sigma_z} ; \qquad \varphi(L)=0,\ \partial_n{\varphi}(0)=0 
\ee
and  $\alpha=\pi$, 
\be\  U'_Z={\  \sigma_z} ; \qquad \varphi(0)=0,\ \partial_n{\varphi}(L)=0 
\ee
 which correspond to  a  Zaremba (mixed) boundary conditions: one boundary is
Dirichlet and the other Neumann. In string theory they correspond to 
strings with one end attached to a 0-dimensional D-brane and the other free
(see figure 5), which also can
be though as attached to a 1-dimensional D-brane \cite{kogan}.

The vacuum energy of Zaremba boundary conditions
\be  E_0=\frac{\pi}{
  L}\left(\frac{1}{48}\right)
\ee
is slightly higher than that of a periodic boundary condition (closed string)
and slightly lower than that of an  antiperiodic boundary condition.

\centerline{\  \epsfxsize7cm \epsfbox{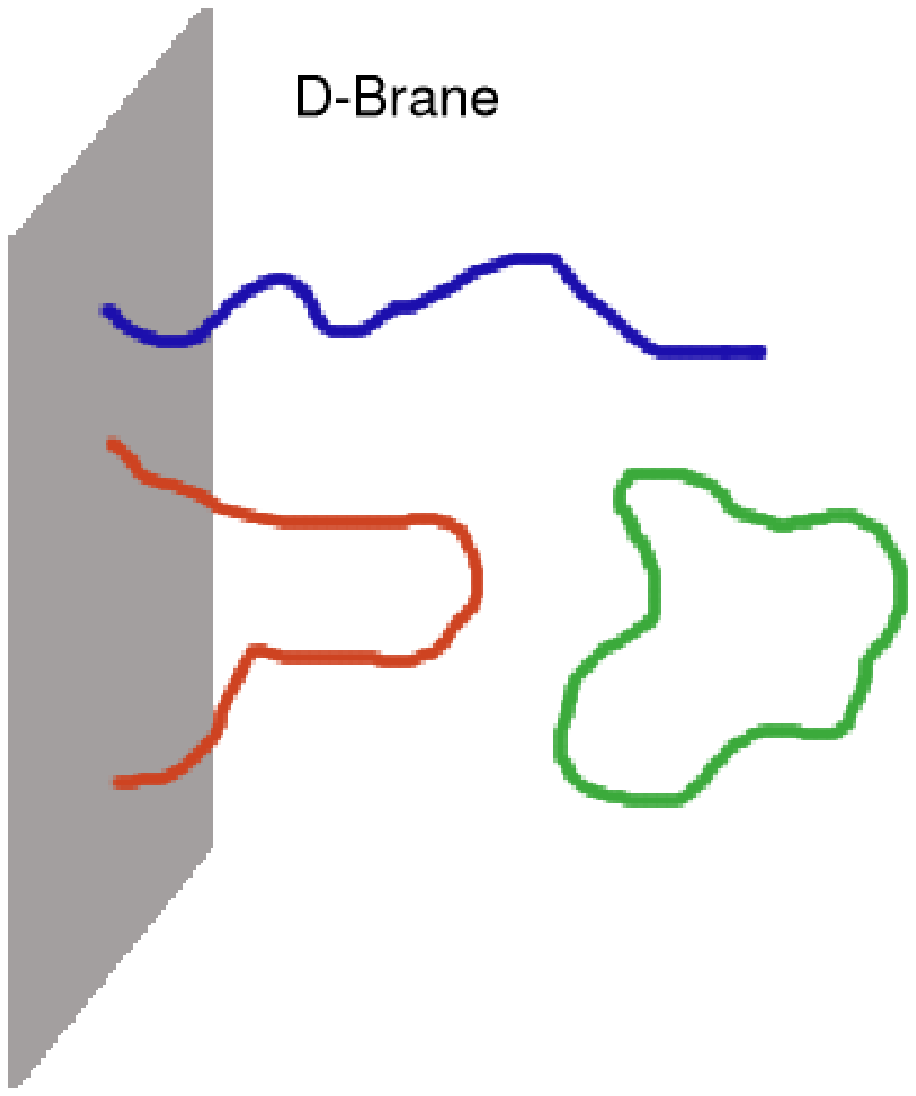}}
\centerline{\small {\bf Figure 5.}  Different types of string attachments to a D-brane}
\noindent

Notice that the global analysis on $\CM$ permits to understand a transition 
from a closed string to an open string with either DD, DN, or NN boundary
conditions \cite{kogan}.
However, as remarked above the transitions to DD or DN might involve an infinite
amount of energy depending on the way the limit is obtained.
The regularity of the interpolation (\ref{qp}) involving a topology change
 is a consequence of the
fact that  $U_\alpha \in
\CC_-\cap \CC_+$ 
for any value of $\alpha$.

\section{ Conformal symmetry, renormalization group  and T-duality}
\vspace{.1cm}
 
 

In 1+1 dimensions   the classical theory of massless scalar fields  is formally conformal
invariant. However, the boundary term (\ref{one}) might break this symmetry already at the
classical level. In the quantum theory there is a conformal anomaly which
makes the realization of conformal invariance more involved, even if
the boundary condition is compatible with the symmetry.

In general, the boundary condition breaks conformal invariance and induces a
renormalization group \cite{Affleck,Zuber,Cardy} flow in the space $\CM$ of  boundary conditions ({\it boundary
renormalization group flow}). Indeed,
 boundary conditions can be introduced into the
action by means of a Lagrange multiplier
\be
\int_{\partial \Omega} \lambda \left(\varphi- i \partial_n\varphi- {   U} \left[\varphi+ i \partial_n\varphi
\right]\right).
\ee
 The renormalization
of this boundary term  defines the renormalization group of 
 boundary conditions.

Conformal invariance is only preserved 
at the fixed points of this boundary renormalization group flow.
These fixed points can easily be  identified. Besides Dirichlet,
Neumann and pseudo-periodic boundary conditions which obviously are
conformal invariant, there are conditions like quasi-periodic boundary
conditions (\ref{quasi})
which also preserve the conformal symmetry.  In 1+1 dimensions they exhaust
the whole set of conformal invariant boundary conditions. The topology of this
subset of fixed points is  $\Zm_2\cup S^2\cup S^2$  in the case of a charged scalar fields and 
 $\Zm_{_2}{\spec{n}} S^1$ in the case of neutral fields. In both cases the topology  of the
space of conformal invariant boundary conditions is still non-trivial.

All other boundary conditions flow towards any of these fixed points.
The renormalization group behaviour around the fixed points is governed by
the Casimir energy and  presents different regimes.
{  {  Dirichlet, Neumann and periodic} boundary f{i}xed points are stable
  whereas quasi-periodic and pseudo-periodic fixed points are in general
unstable and marginally unstable, respectively.

For real scalar fields, Dirichlet, Neumann and periodic boundary conditions are the only stable
points and the result holds for any dimension. The implications of this result
in  string theory are well known. 
Periodic boundary conditions, appear as  attractors of systems with
quasi-periodic  and pseudo-periodic conditions which stresses 
the stability of closed string theory vacuum.

For open strings the attractor (stable) points are  standard free strings
(Neumann) and strings attached to D-branes (Dirichlet). Any other  boundary 
condition flow towards one of those fixed points.

In higher dimensions  ($n>1$) the Hamiltonian (\ref{one})
does not preserves conformal invariance even in the massless case $m=0$.
An extra term 
\be 
\frac{n-1}{4 n}\int \sqrt{g}\, R\,  |\phi|^2
\ee
 proportional to the space-time curvature $R$ has to be added to the action. 
Conformal invariance also requires
a similar modification of Neumann condition in order to preserve 
conformal invariance
\be
\partial_n{\varphi} =\frac{n-1}{4\, n}K\, \varphi,
\ee
where $K$ is the extrinsic curvature of the boundary.
Also more interesting boundary renormalization group flows 
arise. In the case  of systems coupled to magnetic fields  (e.g. Russian doll
models)  or with singular local interactions (see e.g.\cite{tsutsui} for a review),
fixed points with cyclic orbits of the boundary renormalization group flow 
can appear \cite{wilson,sierra,aes}.

Finally, T-duality can also be globally defined  in $\CM$. Indeed,
a T-transformation  is defined by the involutive mapping of a theory with a boundary
condition driven by an operator $U$ into another theory driven by
\be
  U_T= -\sigma_2 U \sigma_2 .
\ee
In particular, T transforms Dirichlet boundary conditions into Neumann boundary
conditions and viceversa. Periodic boundary conditions are T-invariant. More generally,
pseudo-periodic field theories $U_\theta $ are transformed into pseudoperiodic 
theories $U_{-\theta}$.
Notice that  in all cases T preserves the  conformal invariant
nature of  the theory.

 








\section*{Acknowledgments}
We thank M. Aguado, P. Blecua, 
J.G. Esteve, F.S.S. da Rosa, L. B. de Carvalho 
F. Falceto, A. Ibort, D. Lange, G. Marmo and G. Sierra
for interesting discussions on closely related subjects. 
This work is partially supported by CICYT (grant FPA2004-02948)
and DGIID-DGA (grant2005-E24/2).


\bigskip

\end{document}